\documentclass[final,4p]{elsarticle}

\usepackage{amssymb}
\usepackage{bm}
\usepackage{amsmath}
\usepackage{color}
\usepackage[normalem]{ulem}

\pdfoutput=1
\usepackage[caption=false]{subfig}

\journal{Solid State Communications}

\begin{document}

\begin{frontmatter}



\title{Spiral ferrimagnetic phases in the two-dimensional Hubbard model}


\author{J. D. Gouveia\corref{cor1}}
\cortext[cor1]{Corresponding author. Tel. +351 227 828 940}
\ead{gouveia@ua.pt}
\author{R. G. Dias}

\address{Departamento de F\'{\i}sica, I3N, Universidade de Aveiro, Campus de Santiago, Portugal}

\begin{abstract}
We address the possibility of spiral ferrimagnetic phases in the mean-field phase diagram of the two-dimensional (2D) Hubbard model. For intermediate values of the interaction $U$ ($6 \lesssim U/t \lesssim 11$) and doping $n$, a spiral ferrimagnetic phase is the most stable phase in the $(n,U)$ phase diagram. Higher values of $U$ lead to a non-spiral ferrimagnetic phase. If phase separation is allowed and the chemical potential $\mu$ replaces the doping $n$ as the independent variable, the $(\mu,U)$ phase diagram displays, in a considerable region, a spiral (for $6 \lesssim U/t \lesssim 11$) and non-spiral (for higher values of $U$) ferrimagnetic phase with fixed particle density, $n=0.5$, reflecting the opening of an energy gap in the mean-field quasi-particle bands.
\end{abstract}


\end{frontmatter}


\section{Introduction}
\label{}

The 2D Hubbard model remains the most important open theoretical problem in the field of the strongly correlated electronic systems, despite all efforts fuelled by the advent of the high-$T_c$ superconductivity.\cite{Bednorz1986,RevModPhys.70.897} At half-filling, the spin dynamics of the 2D Hubbard model is described by the Heisenberg antiferromagnetic exchange term.\cite{Dias1992} Away from half-filling, the movement of holes through the spin background generates additional spin mixing. The competition between the Heisenberg exchange and the spin configuration mixing generated by hole hopping in the 2D Hubbard model is still far from understood.\cite{Gebhard1997,Ogata1990,Nagaoka1965} In particular, there is no consensus regarding the ground state magnetic phase diagram of the 2D Hubbard model and different authors obtain different mean-field (MF) phase diagrams depending on the magnetic phases allowed.\cite{Marder2000} Traditionally, one considered ferromagnetism, antiferromagnetism and paramagnetism phases.\cite{PhysRev.142.350,J.Dorantes-Davila1083,E.Kaxiras1988,Coppersmith1989,A.Richter1978} The complexity of the MF phase diagram was increased with the introduction of spiral phases\cite{S.Sarker1991}, which appear between the "usual" magnetic phases in the diagram. This complexity was further increased by the consideration of spatial phase separation.\cite{Langmann2007,P.A.Igoshev2010,Schumacher1983}

\begin{figure}[t]
\centering
\includegraphics[height=4cm]{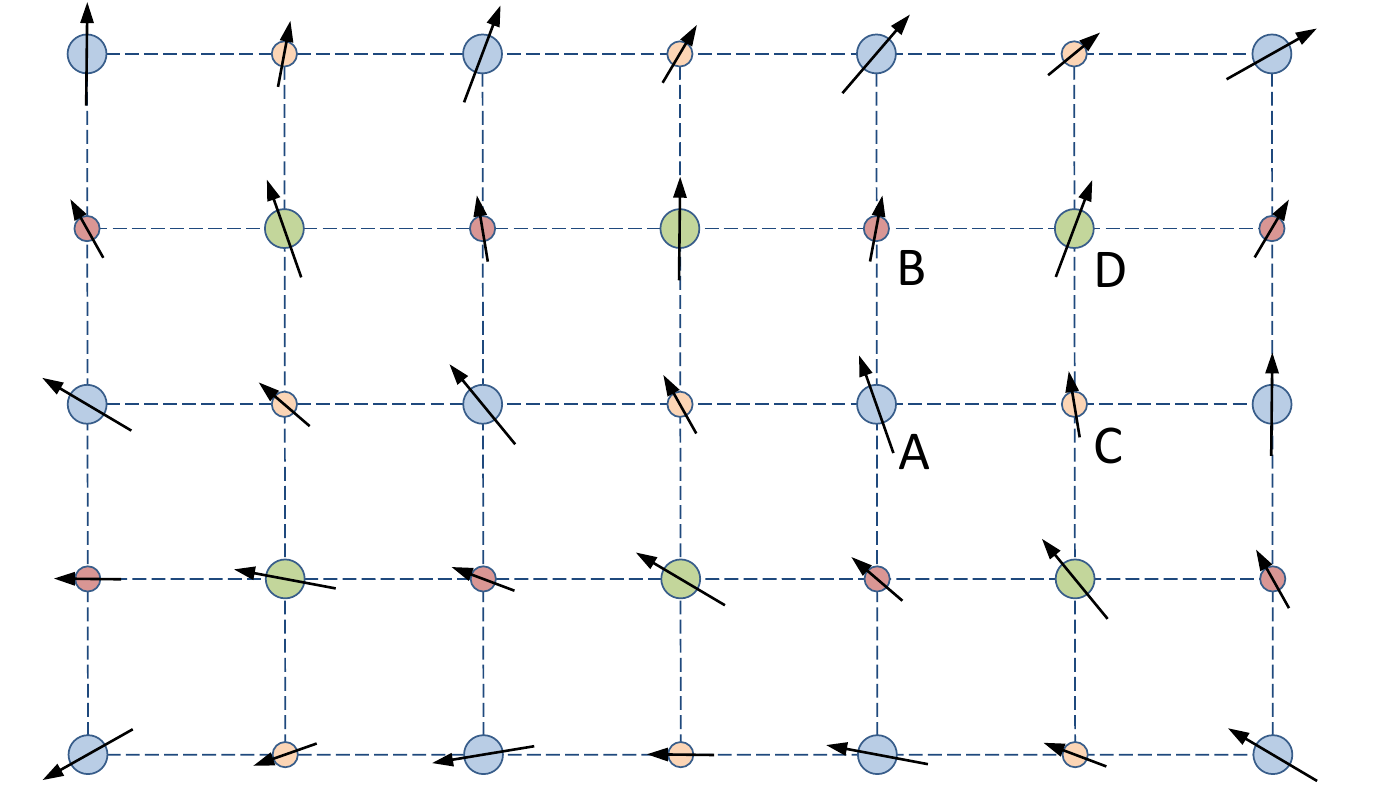}
\caption{The 2D lattice and its four sublattices A, B, C and D. We consider two situations: (i) $m_A = m_D = m_1$ and $m_B = m_C = m_2$; (ii) $m_A = m_C = m_1$ and $m_B = m_D = m_2$.}
\label{fig-Lieb_lattice}
\end{figure}

In this paper, we extend the results above mentioned, by introducing the possibility of a spiral ferrimagnetic phase, that is, a ferrimagnetic phase such that the orientation of magnetic moments changes along the lattice (see Fig. \ref{fig-Lieb_lattice}). More precisely, we study the 2D Hubbard model using the Hartree-Fock approximation in a square lattice decomposing the lattice in four square sublattices (A, B, C and D as in Fig. \ref{fig-Lieb_lattice}) and allowing different amplitudes for magnetizations of the spiral phases in the sublattices. Note that, even under the MF approximation, when four sublattices are considered, it is not possible to obtain the analytical form of the spectra of the 2D Hubbard model. Our MF approach to the 2D Hubbard model follows that of Dzierzawa and Singh.\cite{Dzierzawa1992,A.Singh1992}

\section{Calculations}

Introducing a different creation operator in each sublattice,  $A^\dagger$, $B^\dagger$, $C^\dagger$ and $D^\dagger$, the tight-binding term of the Hubbard Hamiltonian, is
\begin{eqnarray}
H_t = & \sum\limits_{x,y} & A_{x,y}^\dag B_{x,y}   + A_{x,y}^\dag C_{x,y}   \nonumber \\
      &                 + & B_{x,y}^\dag D_{x,y}   + C_{x,y}^\dag D_{x,y}   \nonumber \\
      &                 + & A_{x,y}^\dag B_{x,y-1} + A_{x,y}^\dag C_{x-1,y} \nonumber \\
      &                 + & B_{x,y}^\dag D_{x-1,y} + C_{x,y}^\dag D_{x,y-1} + H.c.,
\label{Eq-Ht_realspace}
\end{eqnarray}
where we set the hopping constant equal to 1.

We consider for now only the sublattice A (we add the other sublattice terms later on). The interaction term of the Hubbard Hamiltonian is, as usual, $H_U = U \sum\limits_r {A_{r\uparrow}^\dag A_{r\uparrow} A_{r\downarrow}^\dag A_{r\downarrow}}$. We assume that the magnetic moments align in the $x$-$y$ plane, so that $\langle S_z \rangle = 0$ and the Hartree term becomes $\frac{U}{4} \sum {  \langle n \rangle^2 }$, where $\langle n \rangle$ is the density of electrons on each sublattice (here assumed to be the same on all of them).

\begin{figure*}[ht]
\centering
\subfloat[$(n,U)$ phase diagram]{\label{fig-Lx100Ly100}\includegraphics[width=.32 \textwidth]{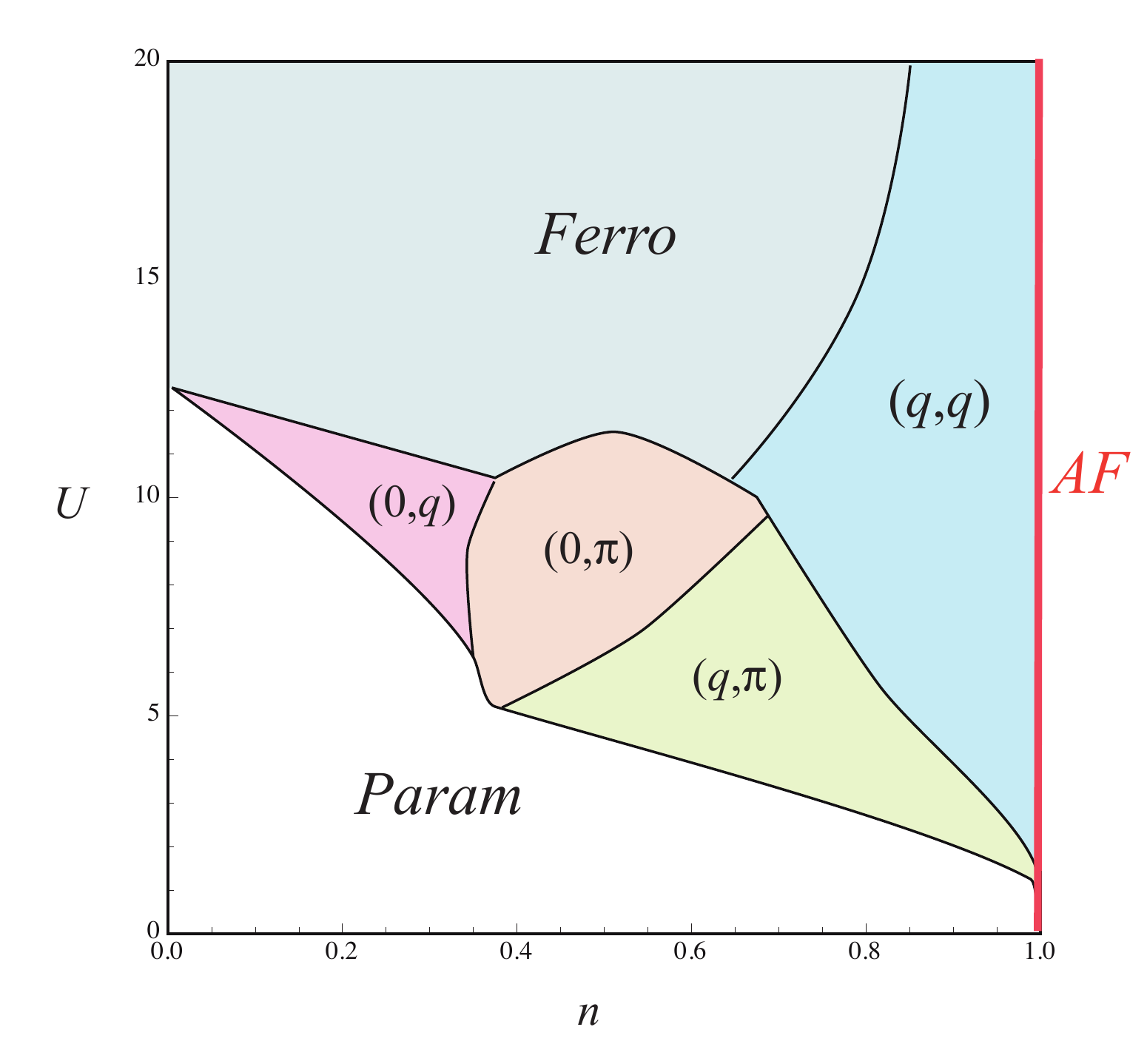}}
\subfloat[$(n,U)$ phase diagram with phase separation]{\label{fig-Lx100Ly100_PS}\includegraphics[width=.32 \textwidth]{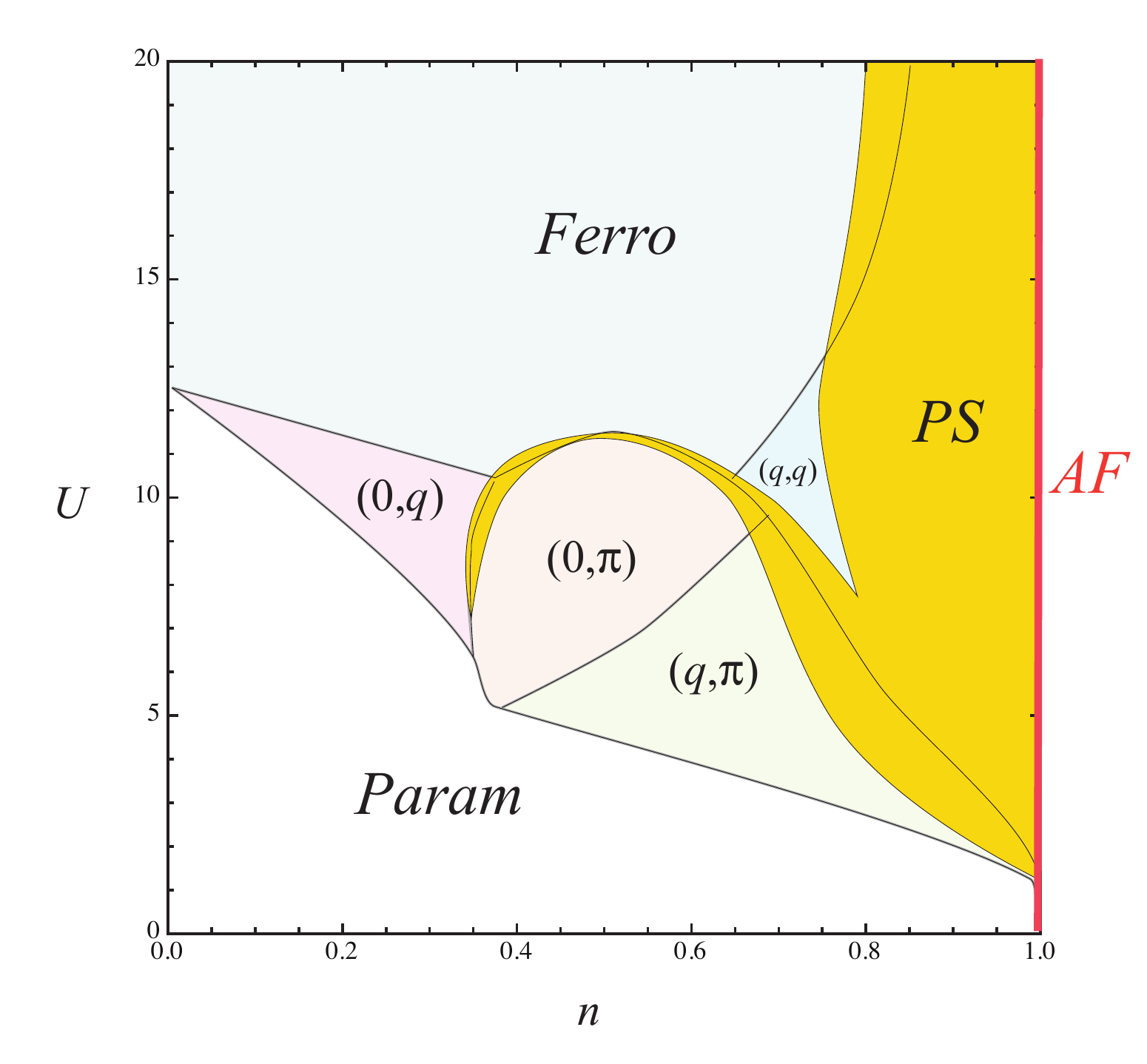}}
\subfloat[$(U,\mu)$ phase diagram with phase separation]{\label{fig-Lx100Ly100_mudiag}\includegraphics[width=.31 \textwidth]{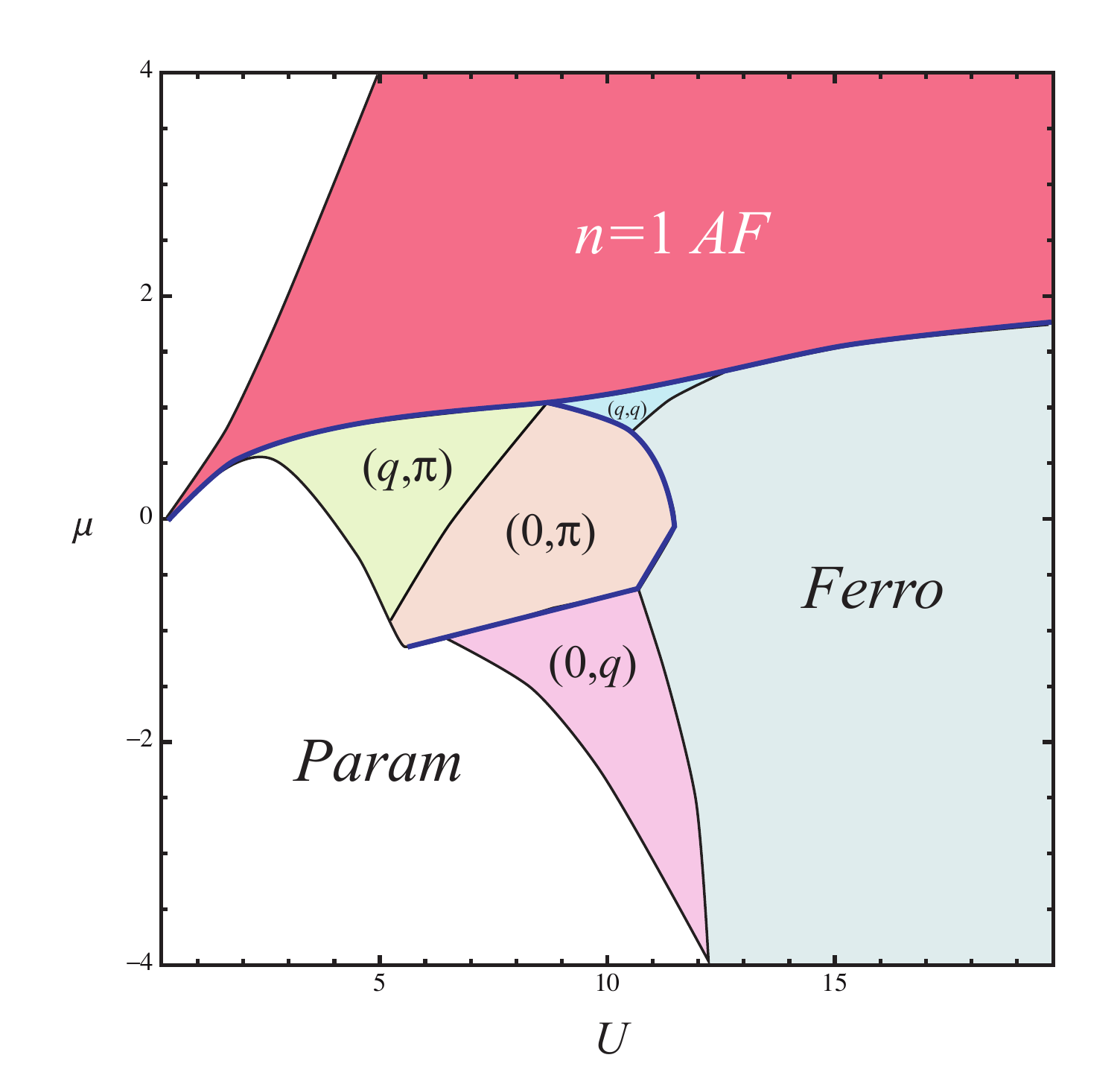}}
\\
\subfloat[$m(n,U)$]{\label{fig-m2D}\includegraphics[width=.2 \textwidth]{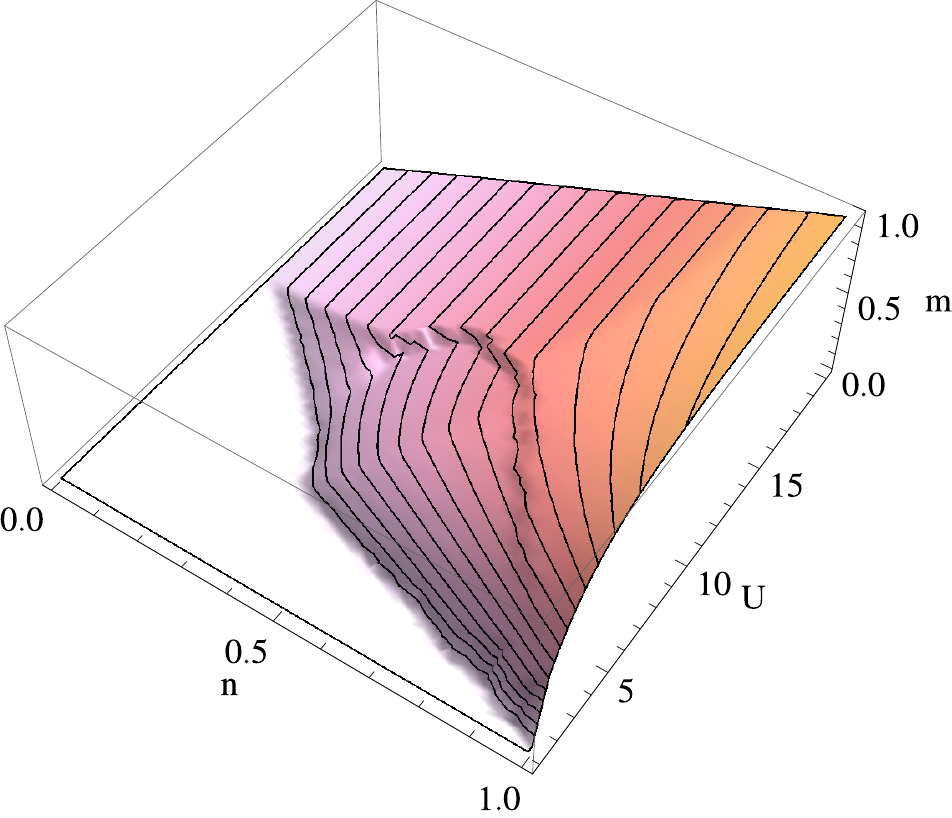}}
\subfloat[$q_x(n,U)$]{\label{fig-Qx2D}\includegraphics[width=.2 \textwidth]{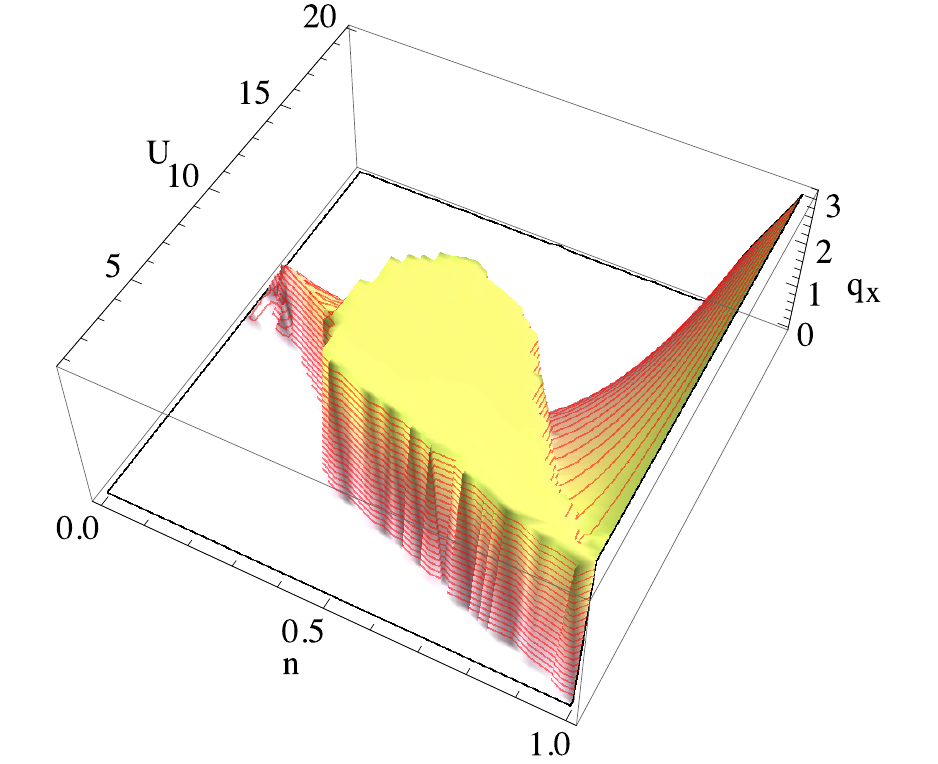}}
\subfloat[$q_y(n,U)$]{\label{fig-Qy2D}\includegraphics[width=.2 \textwidth]{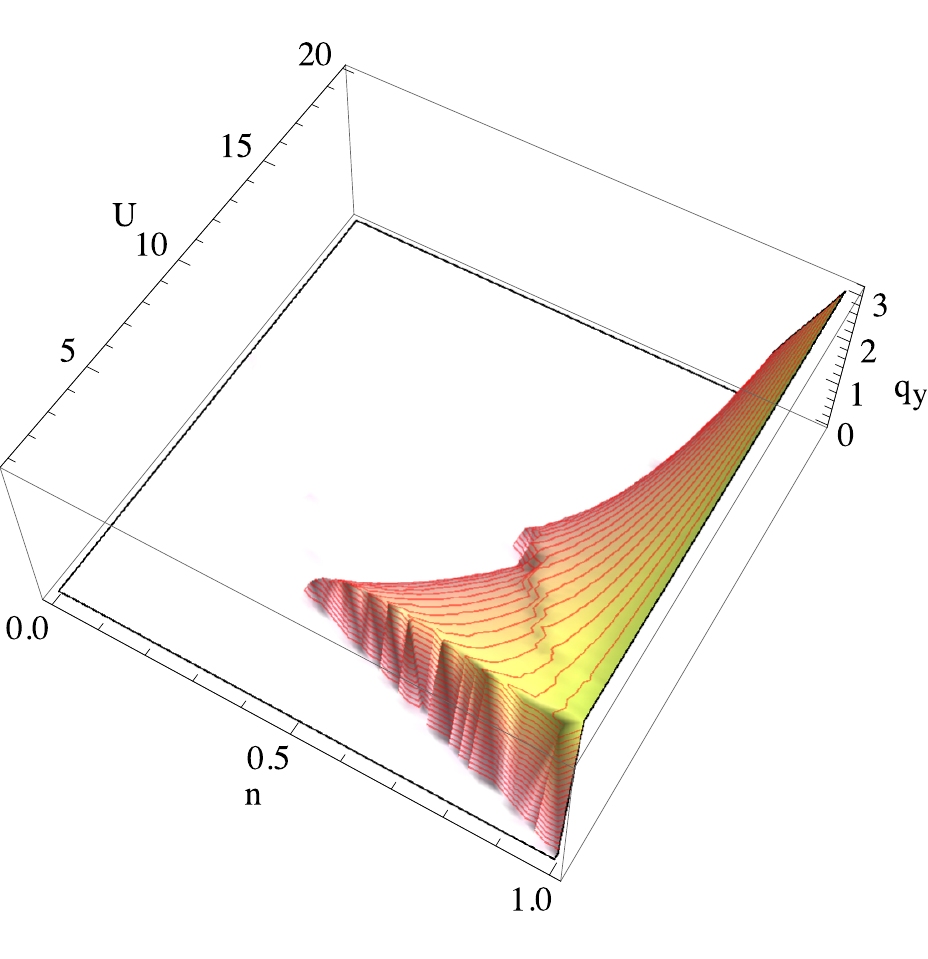}}
\subfloat[$E(n,U)$]{\label{fig-E2D}\includegraphics[width=.2 \textwidth]{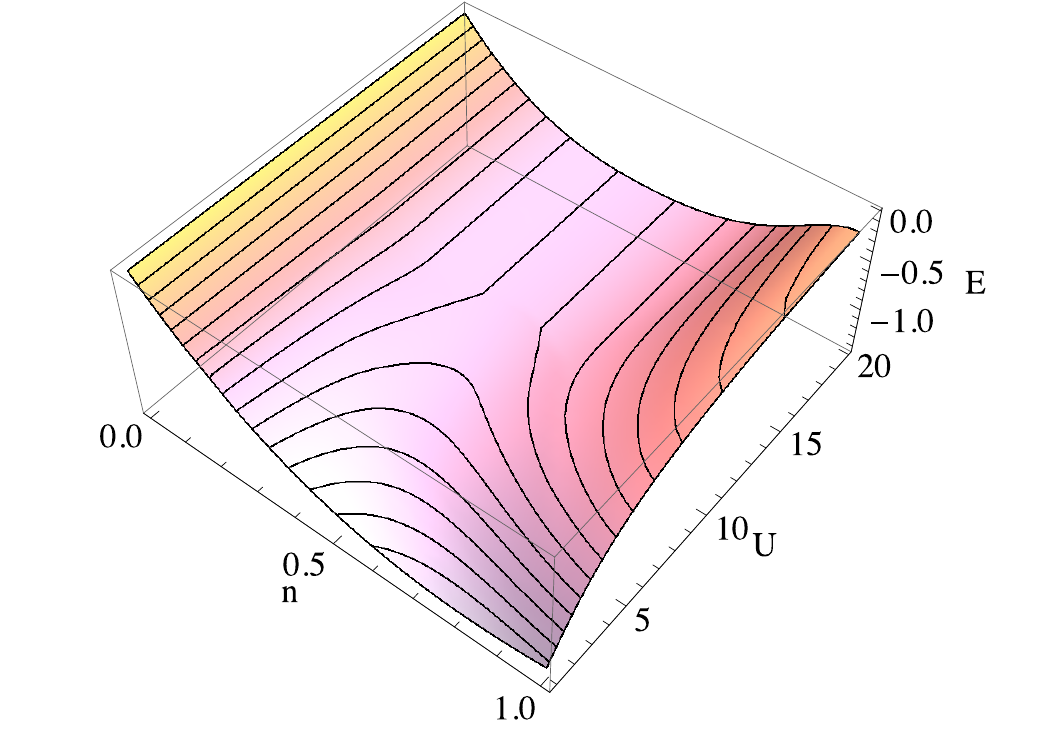}}
\subfloat[$\mu(n,U)$]{\label{fig-mu2D}\includegraphics[width=.2 \textwidth]{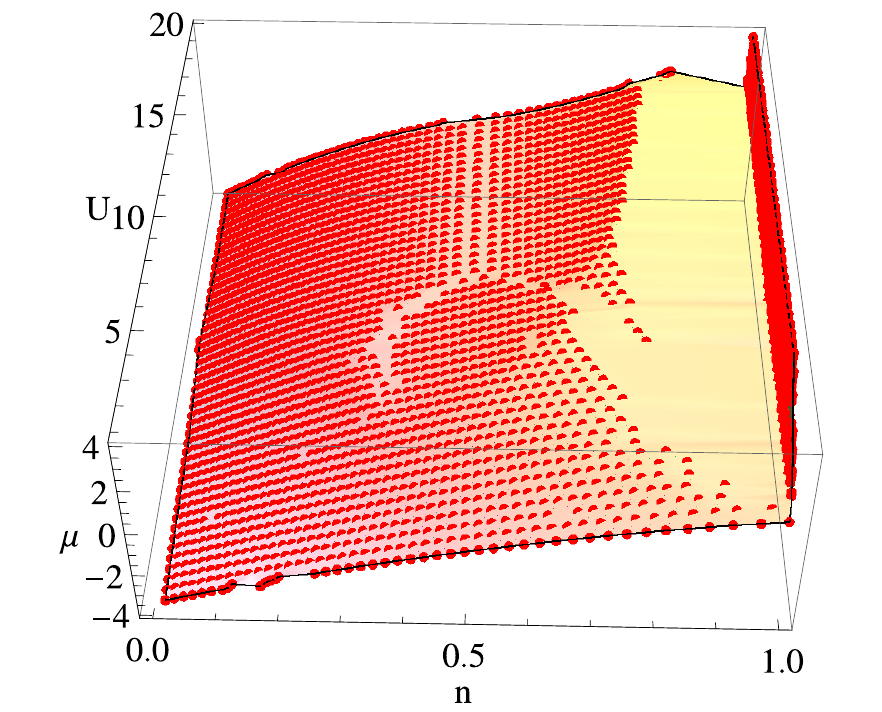}}
\caption{(a) Mean-field  phase diagram for the usual 2D Hubbard model:  The system  displays antiferromagnetism (AF), ferromagnetism (F), paramagnetism (P) or spiral phases ($q \neq 0,\pi$). The antiferromagnetic state $\vec{q} = (\pi,\pi)$ only occurs for $n=1$ (half-filling). (b) $(n,U)$ and (c) $(U,\mu)$ mean-field phase diagrams for the 2D Hubbard model, allowing for phase separation (yellow region). (d) $m$, (e) $q_x$, (f) $q_y$, (g) $E_{MF}$ and (h) $\mu$ as functions of the doping $n$ and Coulomb interaction $U$, for the $100 \times 100$ 2D Hubbard model.}
\label{fig:usualhubbard}
\end{figure*}

The Fock term includes averages like $\langle A_\uparrow^\dag A_\downarrow \rangle = \langle S_A^+ \rangle = \langle S_{Ax} + iS_{Ay} \rangle$, whose values depend on the magnetic phase. Let us assume the average spin in the sublattice A is
\begin{equation}
\langle \vec{S}_{\vec{r}_A} \rangle = \frac{m_A}{2} [\cos ( \vec{q}\cdot\vec{r}_A ) , \sin ( \vec{q}\cdot\vec{r}_A ) , 0 ].
\end{equation}
The vector $\vec{q} = (q_x,q_y)$  defines the magnetic phase of the system. In $\vec{k}$-space we have
\begin{equation}
\langle S_{A\vec{k}}^+ \rangle = \frac{1}{\sqrt{L}} \sum\limits_{\vec{k}'} \langle A_{\vec{k}',\uparrow}^\dag A_{\vec{k}'-\vec{k},\downarrow} \rangle = \frac{m_A\sqrt{L_{u.c.}}}{2} \delta_{\vec{k},-\vec{q}},
\label{Eq-Sak}
\end{equation}
where $L_{u.c.}$ is the number of unit cells, which gives
\begin{equation}
\langle A_{\vec{k},\uparrow}^\dag A_{\vec{k}+\vec{q},\downarrow} \rangle = \frac{m_A}{2},
\end{equation}
while all the other mean values in the summation of Eq. \ref{Eq-Sak} vanish. The Fock term in Fourier space is
\begin{equation}
-\frac{mU}{2} \sum\limits_{\vec{k}} \left( A_{\vec{k}+\vec{q},\downarrow}^\dag A_{\vec{k},\uparrow} + A_{\vec{k},\uparrow}^\dag A_{\vec{k}+\vec{q},\downarrow} \right) + \frac{UL_{u.c.}}{4} m_A^2 .
\end{equation}
Adding the tight-binding, Hartree and Fock terms, the Hamiltonian $H_{MF}$ reads, in the $\{ A_{\vec{k}},B_{\vec{k}},C_{\vec{k}},D_{\vec{k}},A_{\vec{k}+\vec{q}},B_{\vec{k}+\vec{q}},C_{\vec{k}+\vec{q}},D_{\vec{k}+\vec{q}} \}$ basis,
\begin{equation}
\left(
\begin{array}{cc}
H_t(\vec{k}) & H_m \\
H_m^\dag     & H_t(\vec{k}+2\vec{q})
\end{array}
\right),
\end{equation}
plus the diagonal term
\begin{equation}
\frac{UL_{u.c.}}{4} (m_A^2+m_B^2+m_C^2+m_D^2) + \frac{UL\langle n \rangle^2}{2}.
\end{equation}
Here, $H_t(\vec{k})$ is the tight-binding term (Eq. \ref{Eq-Ht_realspace}) of the Hamiltonian in $\vec{k}$-space,
\begin{equation}
\left(
\begin{array}{cccc}
0            & 1+e^{ik_y}  & 1+e^{ik_x}  & 0 \\
1+e^{-ik_y}  & 0           & 0           & 1+e^{ik_x} \\
1+e^{-ik_x}  & 0           & 0           & 1+e^{ik_y} \\
0            & 1+e^{-ik_x} & 1+e^{-ik_y} & 0 \\
\end{array}
\right),
\end{equation}
$H_m $ is the diagonal matrix, $H_m = \text{diag}(\Delta_A,\Delta_B,\Delta_C,\Delta_D)$ with
\begin{equation}
\begin{array}{ll}
\Delta_A = -\dfrac{Um_A}{2}, &
\Delta_B = -\dfrac{Um_B}{2}e^{iq_y}, \\
\Delta_C = -\dfrac{Um_C}{2}e^{iq_x}, &
\Delta_D = -\dfrac{Um_D}{2}e^{iq_x+iq_y}.
\end{array}
\end{equation}

\begin{figure*}[ht]
\centering
\subfloat[$(n,U)$ phase diagram]{\label{fig-Lx100Ly100_dimer}\includegraphics[width=.4 \textwidth]{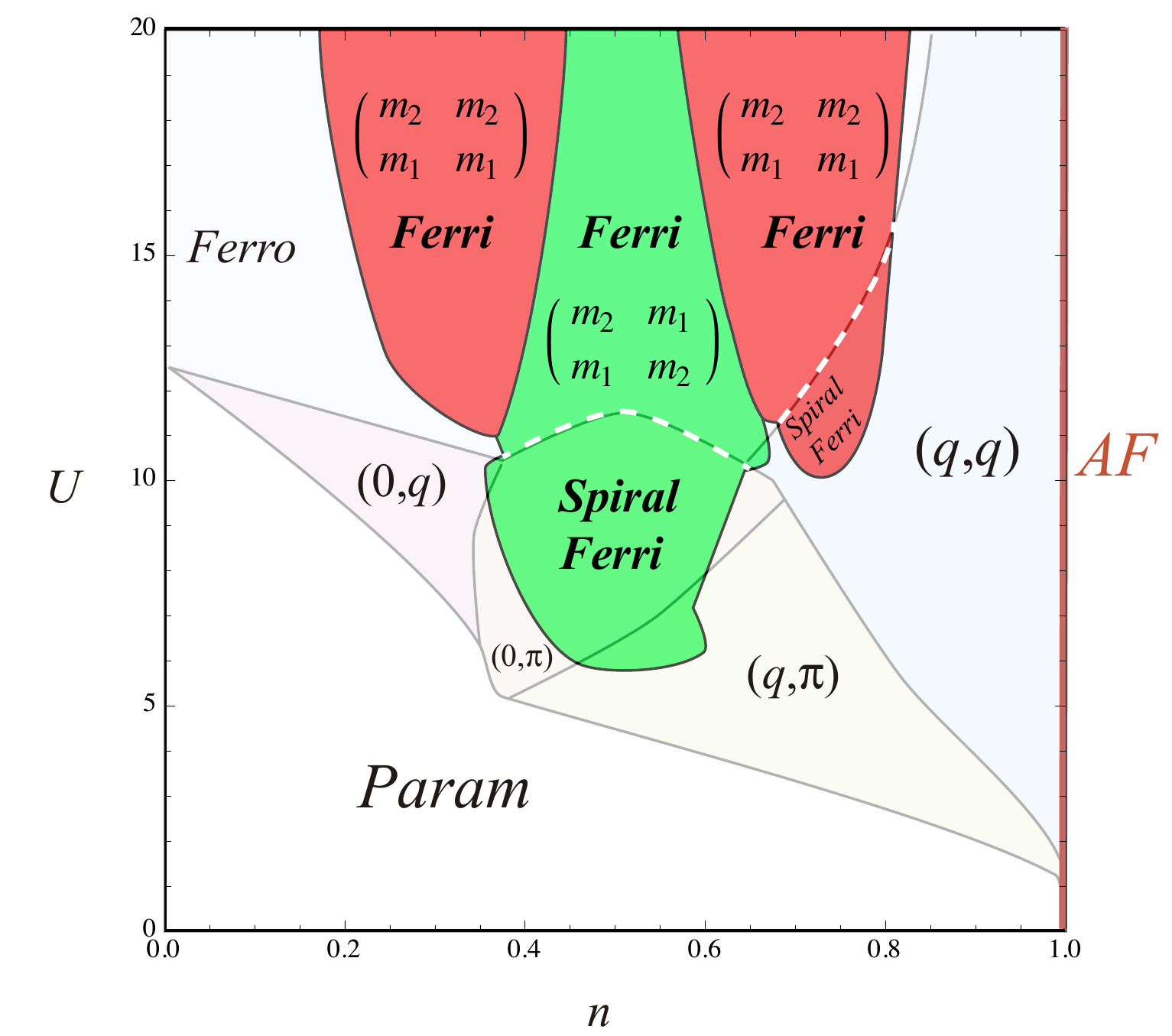}}
\subfloat[$(U,\mu)$ phase diagram with phase separation]{\label{fig-Lx100Ly100_mu_dimer}\includegraphics[width=.385 \textwidth]{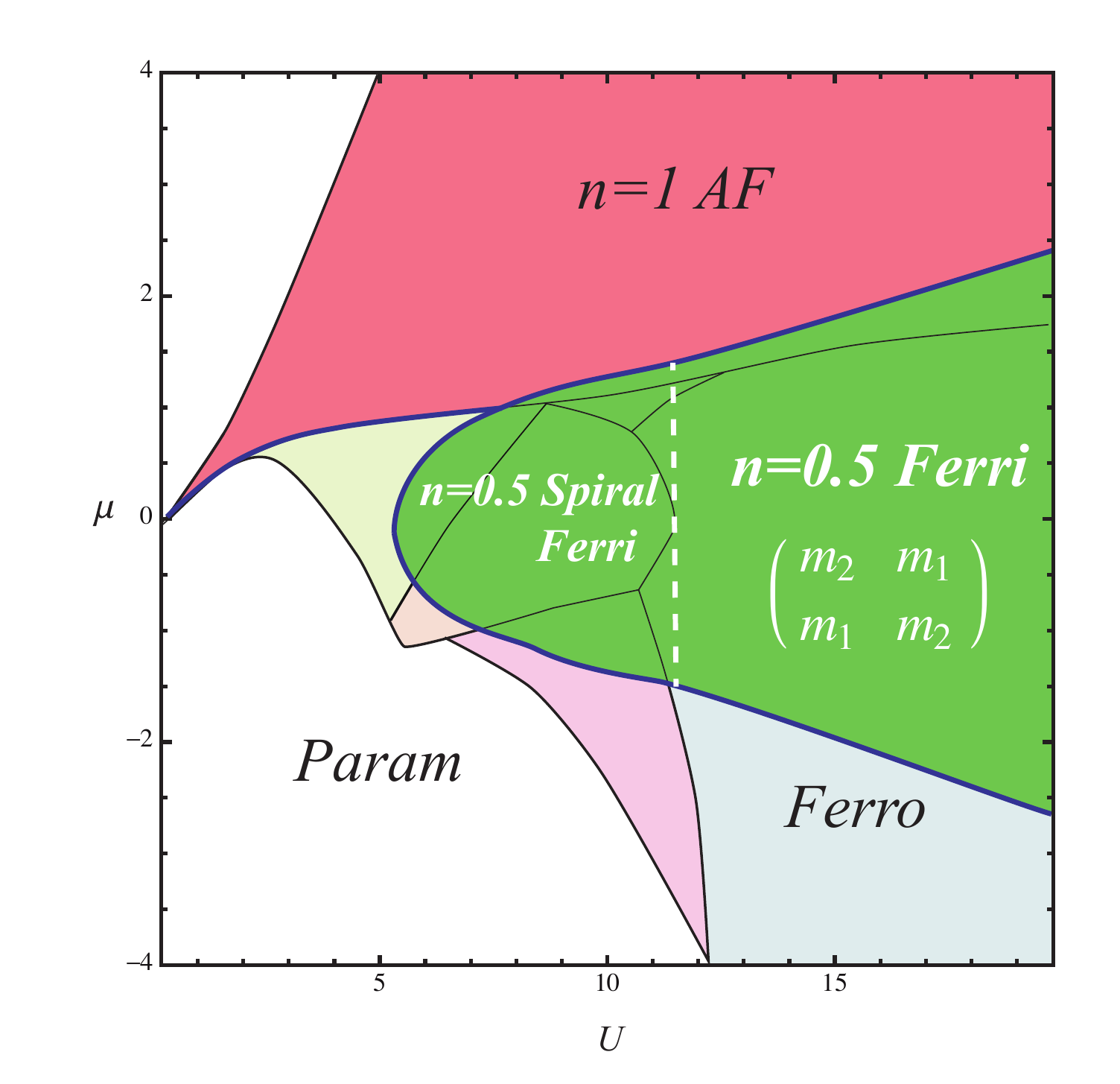}}
\\
\subfloat[$m_1,m_2$ $(i) $]{\label{fig-m1m2m2m1}\includegraphics[width=.25 \textwidth]{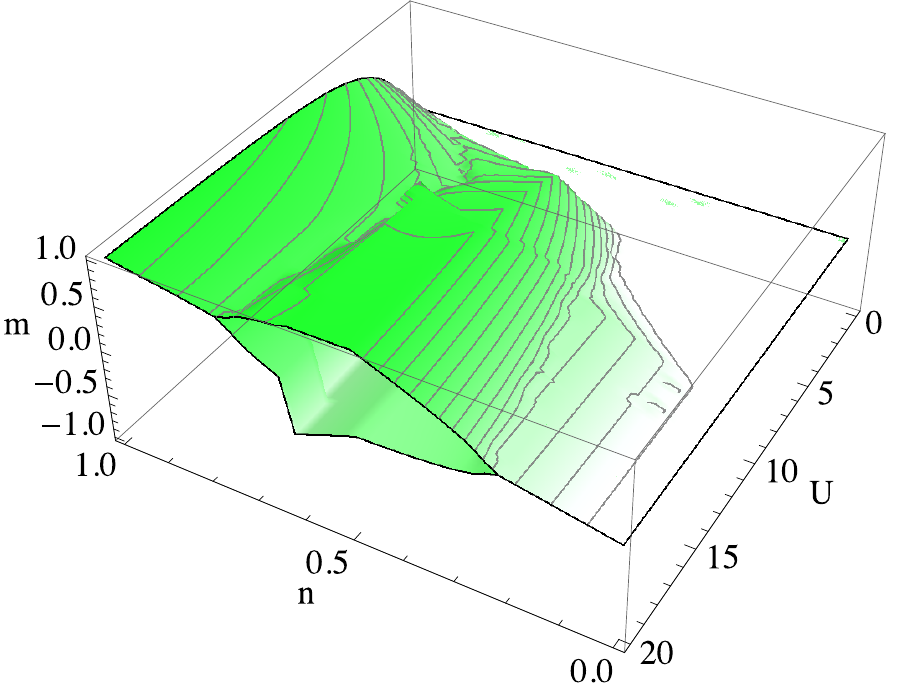}}
\subfloat[$m_1,m_2$ $(ii)$]{\label{fig-m1m1m2m2}\includegraphics[width=.25 \textwidth]{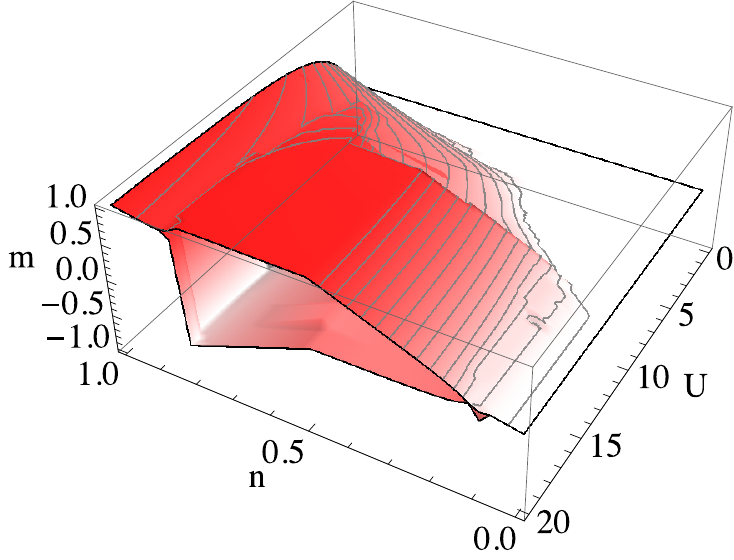}}
\subfloat[$E(n,U)$]{\label{fig-compare_m_configurations}\includegraphics[width=.2 \textwidth]{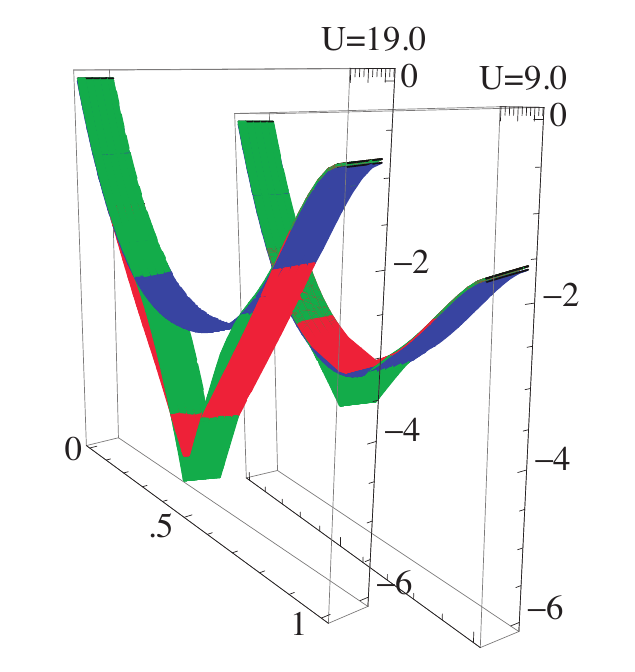}}
\subfloat[$\mu(n,U)$]{\label{fig-mudimer}\includegraphics[width=.3 \textwidth]{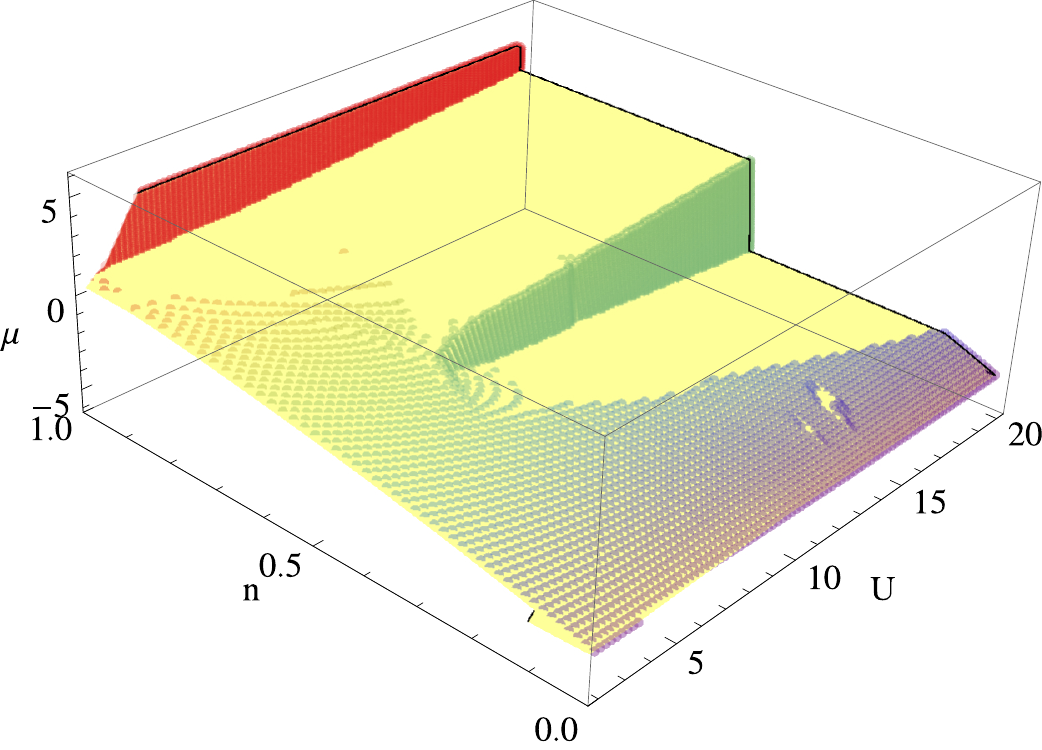}}
\caption{Top figures: (a) Mean-field  phase diagram for the 2D Hubbard model with sublattices with independent magnetization amplitudes. In the central (green) region of the phase diagram, configuration (i) is the one that minimizes the MF energy, while configuration (ii) is the most stable in the red regions. (b) $(U,\mu)$ mean-field phase diagram with phase separation occurring on the borders of the green and red regions (thick solid lines). This phase diagram displays spiral and non-spiral ferrimagnetic phases (green regions) with fixed particle density, $n=0.5$, reflecting the opening of an energy gap in the MF quasi-particle bands. Bottom figures: $m_1(n,U)$ and $m_2(n,U)$ for the MF 2D Hubbard model with two sublattices in (c) configuration (i) and (d) configuration (ii). (e) Ground state MF energy for usual 2D (blue), case (i) (green) and case (ii) (red) for $U=9$ and $U=19$. (f) $\mu$ as a function of the doping $n$ and Coulomb interaction $U$. All figures represent numerical results obtained for the $100 \times 100$ 2D Hubbard model with two sublattices with different magnetization amplitudes.}
 \label{fig:dimerhubbard}
\end{figure*}

\section{Results and discussion}

By setting $m_A=m_B=m_C=m_D=m$, we recover the MF magnetic phase diagram of the usual 2D Hubbard model, consistent with the ones obtained by several authors\cite{Dzierzawa1992,S.Sarker1991,P.A.Igoshev2010} for zero temperature, as presented in Fig. \ref{fig-Lx100Ly100}. In order to obtain such a diagram, one minimizes either the MF energy $E_{MF}$ using the electronic density $n$ as an independent variable, or the thermodynamic potential $\Omega_{MF}$ using the chemical potential $\mu$, with respect to the site magnetization amplitude $m$ and the order parameter $\vec{q} = (q_x,q_y)$. These parameters define the magnetic phase of the system.

A solution with $m=0$ is paramagnetic and is usually $\vec{q}$-degenerate, while solutions for $m\neq 0$ are in general unique. In the latter case, the wave vector $\vec{q}$ specifies the type of magnetic ordering. For instance, $\vec{q} = (0,0)$ for the ferromagnetic phase, $\vec{q} = (\pi,\pi)$ for the antiferromagnetic phase and all other choices for spiral phases. In the example shown in Fig. \ref{fig-Lieb_lattice}, we have $q_x = \pi/18$ and $q_y = \pi/6$. Additionally, in the same example, the magnetization amplitudes (denoted by the size of the arrows) are $m_A=m_D=m_1$ and $m_B=m_C=m_2<m_1$. Comparing, for each pair $(n,U)$ or $(U,\mu)$, the data obtained for $m$ (Fig. \ref{fig-m2D}), $q_x$ (Fig. \ref{fig-Qx2D}) and $q_y$ (Fig. \ref{fig-Qy2D}), the MF magnetic phase diagram displayed in Fig. \ref{fig-Lx100Ly100} ensues. For some values of $\mu$, there is more than one pair $( \vec{q},m )$ which minimizes the thermodynamic potential. In those cases, a first-order phase transition in the order parameters occurs. When using $n$ as a basic variable (and posteriorly calculating $\mu = \partial E / \partial n \approx \Delta E/\Delta n$ using the data in Fig. \ref{fig-E2D}), $n$ seems to be multiply defined for some values of $\mu$, which implies instability (e. g. of the spiral phase for $U=15$). The use of $\mu$ as a basic variable solves this ambiguity and leads to plateaus in the chemical potential $\mu(n,U)$ in the regions where phase separation (PS) occurs (see Fig. \ref{fig-mu2D}). In each PS region of the diagram, two spatially separated phases occur: the ones immediately to the left and to the right of the PS region in question (see Fig. \ref{fig-Lx100Ly100_PS}). The two phases have different electronic densities, such that the electronic density of the whole system amounts to $n$. In Fig. \ref{fig-Lx100Ly100_mudiag}, we show the same phase diagram as in Fig. \ref{fig-Lx100Ly100}, but using $\mu$ as the independent variable. The colors of corresponding regions are the same for easier reading. The thick solid line indicates a discontinuity in $n$.

In this work, the magnetic phase diagram for the Hubbard 2D model comprising four sublattices is obtained by finding the magnetization amplitudes $(m_A,m_B,m_C,m_D)$ and the vector $\vec{q}$ which minimize the energy. We consider two situations: (i) $m_A = m_D = m_1$ and $m_B = m_C = m_2$; (ii) $m_A = m_C = m_1$ and $m_B = m_D = m_2$.

The ground state magnetization amplitude of the usual 2D Hubbard model is proportional to $n$ for each value of $U$ in the ferromagnetic phase (see Fig. \ref{fig-m2D}). When a spiral ferrimagnetic phase is allowed, it was found that, near zero filling ($n=0$) and half-filling ($n=1$), the ground state magnetization remains the same as in the usual 2D case (see Figs. \ref{fig-m1m2m2m1} and \ref{fig-m1m1m2m2}). This means that in these regions, the ground state magnetization is still constant throughout the whole lattice. However, as one moves to intermediate $n$, one finds that $m_1$ and $m_2$ become distinct, as shown in Figs. \ref{fig-m1m2m2m1} and \ref{fig-m1m1m2m2} for cases (i) and (ii) respectively, where $m_1$ and $m_2$ are displayed as a function of $n$ and $U$. These figures show two sheets reflecting the separation of the magnetization amplitudes. The colors green for case (i) and red for (ii) are used on all plots of Fig. \ref{fig:dimerhubbard}. For intermediate filling, the system is able to lower its energy by adopting different magnetization amplitudes on sublattices $1$ and $2$ in both cases (i) and (ii). This is shown in Fig. \ref{fig-compare_m_configurations} for $U=19$ and $U=9$. Depending on the region of the phase diagram one analyses, configuration (i) or (ii) may have the lowest energy, as shown in Fig. \ref{fig-Lx100Ly100_dimer}. In this figure, we added another layer on top of the usual 2D MF magnetic phase diagram, showing which of the two-sublattice configurations considered has the lowest energy in the ferrimagnetic region: green for case (i) and red for case (ii). Furthermore, the energy was minimized with respect to $q_x$ and $q_y$, while using the new magnetization values, but it was found that only very small changes in $\vec{q}$ occur, i.e., despite the changes in magnetization amplitudes, the magnetic phases in the diagram remain the same. For this reason, the magnetic phases are shown as being the same as those of the usual 2D model.

The mean-field energy dispersion relation of the usual 2D Hubbard model displays two bands. Electrons occupy the lowest band until half-filling ($n=1$) and then proceed to occupying the higher band. As can be seen in Fig. \ref{fig-mu2D}, the fermionic density increases with the chemical potential until the phase separation region is reached. In this region, the chemical potential is constant despite any increase in the number of particles, up to half-filling. At this point, any increase in $n$ induces a jump in the value of $\mu$, equal to the energy separation between the two energy bands (called the energy gap). As the plot in Fig. \ref{fig-mu2D} only goes up to half-filling, we see $\mu$ increasing smoothly until it reaches the phase separation region, followed by a plateau and a jump at $n=1$. In both cases studied in this work, with the lattice divided into two sublattices, the energy bands open a gap at quarter filling ($n=0.5$), as shown in Fig. \ref{fig-mudimer}. Another gap appears at three quarter filling ($n=0.75$), but only the phases with $n \leq 1$ are shown in Fig. \ref{fig-Lx100Ly100_mu_dimer}. The plot in Fig. \ref{fig-Lx100Ly100_mu_dimer} is again the same as Fig. \ref{fig-Lx100Ly100_dimer}, but using $\mu$ as the basic variable. In this diagram, the green region corresponds to $n=0.5$, therefore only configuration (i) for the spiral and non spiral ferrimagnetic phases is present in the phase diagram. The dashed line separates the ferrimagnetic region from the spiral ferrimagnetic one and the thick solid lines denote again discontinuities in $n$.

\section{Conclusion}

Having addressed the possibility of a spiral ferrimagnetic phase in the mean-field phase diagram of the 2D Hubbard model, we conclude that, for intermediate values of the interaction $U$ and doping $n$, the spiral ferrimagnetic phase is the most stable phase in the $(n,U)$ phase diagram. Higher values of $U$ lead to non-spiral ferrimagnetic phases. We emphasize the case of intermediate $n$ and higher $U$, for which the ground state does not appear to be purely ferromagnetic, contrasting with results by other authors.\cite{P.A.Igoshev2010} Additionally, allowing phase separation and replacing $n$ by $\mu$ as the independent variable, the $(\mu,U)$ phase diagram displays, in a considerable region, spiral (for intermediate values of $U$) and non-spiral (for higher values of $U$) ferrimagnetic phases with fixed particle density, $n=0.5$, reflecting the opening of an energy gap in the mean-field quasi-particle bands. We further note that generalizing the ferrimagnetic phase to cases where more than two different magnetization amplitudes are allowed should lead to even more stable ferrimagnetic phases in certain regions of the phase diagram. Preliminary results with three different amplitudes are consistent with this conjecture. All these results provide strong evidence on the stability of the spiral ferrimagnetic phase in the mean field magnetic phase diagram of the 2D Hubbard model.

\section*{Acknowledgements}

R. G. Dias acknowledges the financial support from the Portuguese Science and Technology Foundation (FCT) through the program PEst-C/CTM/LA0025/2013. J. D. Gouveia acknowledges the financial support from the Portuguese Science and Technology Foundation (FCT) through the grant SFRH/BD/73057/2010.





\bibliographystyle{model1a-num-names}
\bibliography{helix1}







\end{document}